\documentclass[fleqn,usenatbib]{mnras}

\usepackage[T1]{fontenc}
\usepackage{ae,aecompl}

\usepackage{graphicx}
\title{Resolving the X-ray emission from the Lyman continuum emitting galaxy Tol 1247-232}

\author[Kaaret, Brorby, Casella, Prestwich ]{
P. Kaaret$^{1}$ \thanks{E-mail: philip-kaaret@uiowa.edu},
M. Brorby$^{1}$,
L. Casella$^{1}$,
A.H. Prestwich$^{2}$
\\
$^{1}$Department of Physics and Astronomy, University of Iowa, Iowa City, IA 52245, USA\\
$^{2}$Harvard-Smithsonian Center for Astrophysics, Cambridge, MA, 02318, USA\\
}

\date{Accepted XXX. Received YYY; in original form ZZZ}

\pubyear{2017}

\begin{document}
\label{firstpage}
\pagerange{\pageref{firstpage}--\pageref{lastpage}}
\maketitle

\begin{abstract}

{\it Chandra} observations of the nearby, Lyman-continuum (LyC) emitting galaxy Tol 1247-232 resolve the X-ray emission and show that it is dominated by a point-like source with a hard spectrum ($\Gamma = 1.6 \pm 0.5$) and a high luminosity ($(9 \pm 2) \times 10^{40} \rm \, erg \, s^{-1}$). Comparison with an earlier {\it XMM-Newton} observation shows flux variation of a factor of 2. Hence the X-ray emission likely arises from an accreting X-ray source: a low-luminosity AGN or one or a few X-ray binaries. The {\it Chandra} X-ray source is similar to the point-like, hard spectrum ($\Gamma = 1.2 \pm 0.2$), high luminosity ($10^{41} \rm \, erg \, s^{-1}$) source seen in Haro 11, which is the only other confirmed LyC-emitting galaxy that has been resolved in X-rays. We discuss the possibility that accreting X-ray sources contribute to LyC escape.

\end{abstract}

\begin{keywords}
galaxies: star formation -- galaxies: individual: Tol 1247-232, Haro 11 -- X-rays: galaxies -- X-rays: binaries
\end{keywords}

%%%%%%%%%%%%%%%%% BODY OF PAPER %%%%%%%%%%%%%%%%%

\section{Introduction}

Between 100 million and 1 billion years after the Big Bang, the intergalactic medium (IGM) changed from being cold and neutral to being warm and ionized.  This reionization of the universe required sources of radiation energetic enough to ionize neutral hydrogen. Massive stars produce copious Lyman radiation and are often considered the most likely sources of reionization \citep{Loeb2010}. However, Lyman radiation is efficiently absorbed by the gas from which stars form and the details of how it escapes from a galaxy are poorly understood. Lyman continuum (LyC) and line emission is absorbed by dust and the Ly$\alpha$ is resonantly scattered by neutral hydrogen. Both reduce the escape fraction below values predicted by simple models of HII regions (Hayes et al.\ 2010). It appears that some source of feedback is required to blow the neutral gas and dust away from the starburst to prevent scattering and allow the Lyman emission to escape \citep{Wofford2013,Orsi2012}. Obvious sources of mechanical power are stellar winds and supernovae ejecta \citep{Tenorio1999,Hayes2010,Heckman2011}.  Another potential source of mechanical power is outflows from accreting compact objects. The power in the outflow from an accreting object is often comparable to the radiative luminosity \citep{Gallo2005,Justham2012}. In some systems, particularly those found in actively star forming galaxies \citep{Kaaret2017}, there is evidence that the mechanical power exceeds the radiative luminosity by large factors \citep{Pakull2010}.

Most searches for LyC radiation from local galaxies have come up empty handed \citep{Leitherer1995,Grimes2007}. There are only three known nearby (closer than 1000~Mpc) galaxies from which LyC emission has been directly detected: Haro~11 \citep{Bergvall2006}, Tololo 1247-232 \citep[][hereafter Tol1247]{Leitet2013}, and Mrk 54 \citep{Leitherer2016}. Remarkably, the two that have been observed in X-rays are very luminous with $L_X \sim 10^{41}$~erg/s in the 0.5--10~keV band for both Tol1247 \citep{Rosa2009} and Haro 11 \citep{Prestwich2015}.

High resolution Chandra imaging of Haro~11 shows diffuse thermal emission ($kT \sim 0.7$~keV) and two bright point sources: one with a luminosity of $\sim 10^{41}$~erg/s and a very hard spectrum ($\Gamma = 1.2 \pm 0.3$) located in a star-forming knot and second with a luminosity of $5 \times 10^{40}$~erg/s and a softer spectrum ($\Gamma \sim 2.2$) located 3 arcseconds away in another star-forming knot \citep{Prestwich2015}. The sources are likely accreting objects. Modelling of the mechanical power from the central star-forming knots shows that the total mechanical power in supernovae plus stellar winds is comparable to the luminosity of the X-ray sources \citep{Prestwich2015}. Thus, if the X-ray sources produce outflows with power comparable to their luminosities, then feedback from accreting X-ray sources may help enable Lyman escape.

We recently obtained {\it Chandra} observations of the nearby LyC-emitting galaxy, Tol1247, that permit us to resolve the X-ray emission on physical scales of 500~pc, smaller than the 10~kpc optical extent of the galaxy.  We describe the observations and analysis in section~\ref{sec:obs} then discuss our results and their interpretation in section~\ref{sec:results}. We adopt a distance to Tol1247 of 207~Mpc and a redshift of 0.048 \citep{Leitherer2016}.

\section{Observations and Analysis}\label{observations}
\label{sec:obs}

\begin{figure}
\centerline{\includegraphics[width=2.75in]{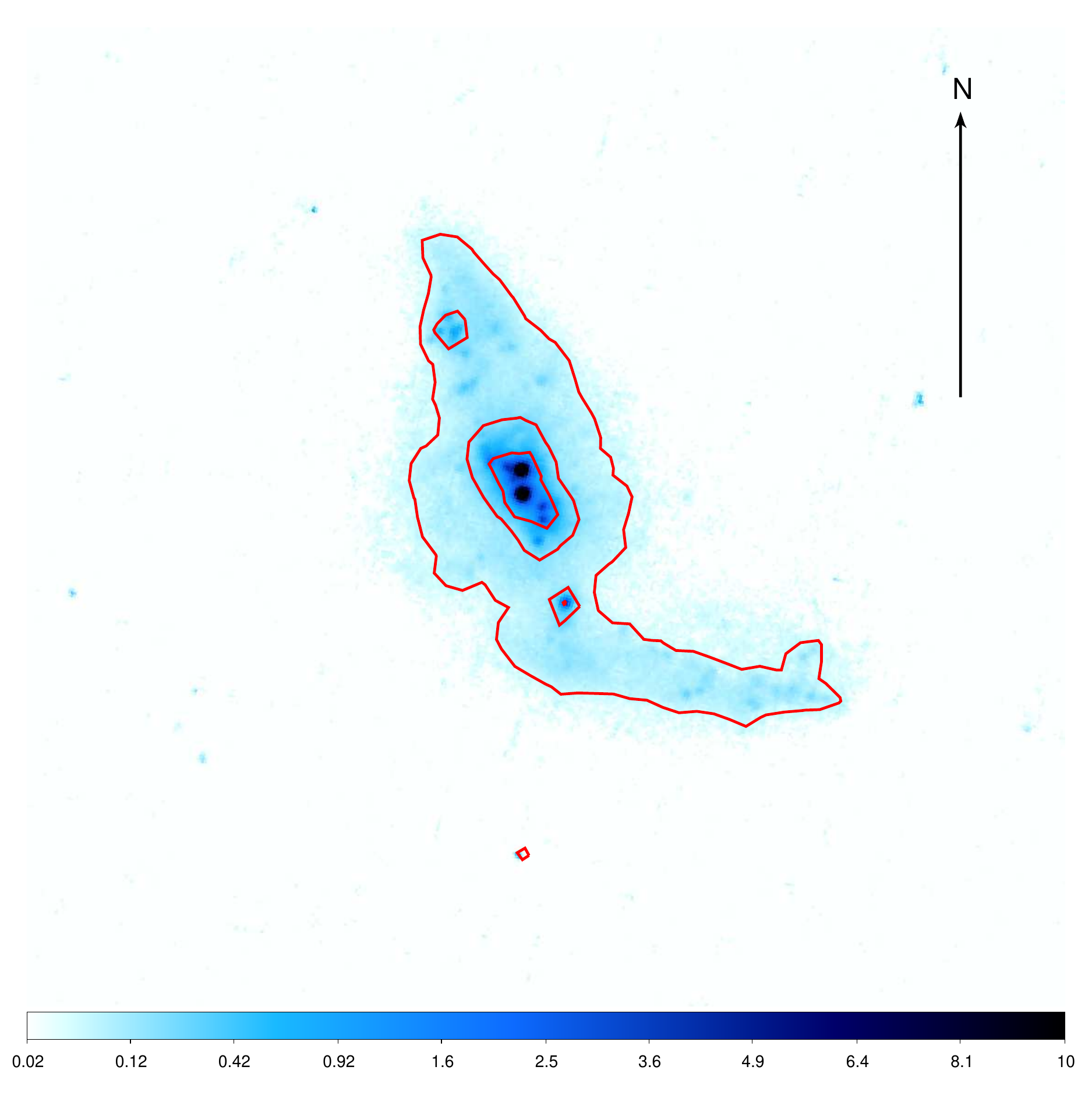}}
\caption{Optical image of Tol 1247-232. An HST image in the F438W filter that shows the optical extent of the irregular galaxy used to define the red contours in Fig.~\ref{xray_images}. The black arrow points North and has a length of $4\arcsec$.}
\label{optical_image}
\end{figure}

Figure~\ref{optical_image} shows a Hubble Space Telescope (HST) image of Tol1247 obtained with the WFC3 using the F438W filter (B-band). Using the Graphical Astronomy and Image Analysis Tool ({\it GAIA}), we aligned the HST image to stars in the 2mass catalogue that has an astrometric accuracy of $0.2\arcsec$ \citep{2mass}. We found 10 coincidences with bright, isolated stars in the HST image and the root mean square deviation between the stellar centroids and catalogue positions was $0.22\arcsec$.

Our approved 30~ks {\it Chandra} observation was divided into two 15~ks observations for operational reasons. The first began at 07:26:01 UTC on 2016-05-13 (ObsID 16971, hereafter observation A) and the second began at 06:52:32 UTC on 2016-05-14 (ObsID 18845, hereafter observation B), about one day later. We used CIAO version 4.8 with data processing version of 10.4.3.1 and CALDB version 4.7.1.

There is strong emission coincident with the galaxy in both observations. Using a source extraction region with a radius of 3$\arcsec$ encompassing the whole galaxy, we found a net count rate in the 0.3--8~keV band of $2.5 \pm 0.4$~c/ks for A and $3.1 \pm 0.5$~c/ks for B. There is no evidence for variability between the two observations, so we chose to combine the two observations for further analysis.
% We also extracted spectra in the 0.3--8~keV band and fitted each with an absorbed powerlaw with fixed absorption and found no evidence for variability of the spectral index or normalization.

\begin{figure}
\centerline{\includegraphics[width=2.75in]{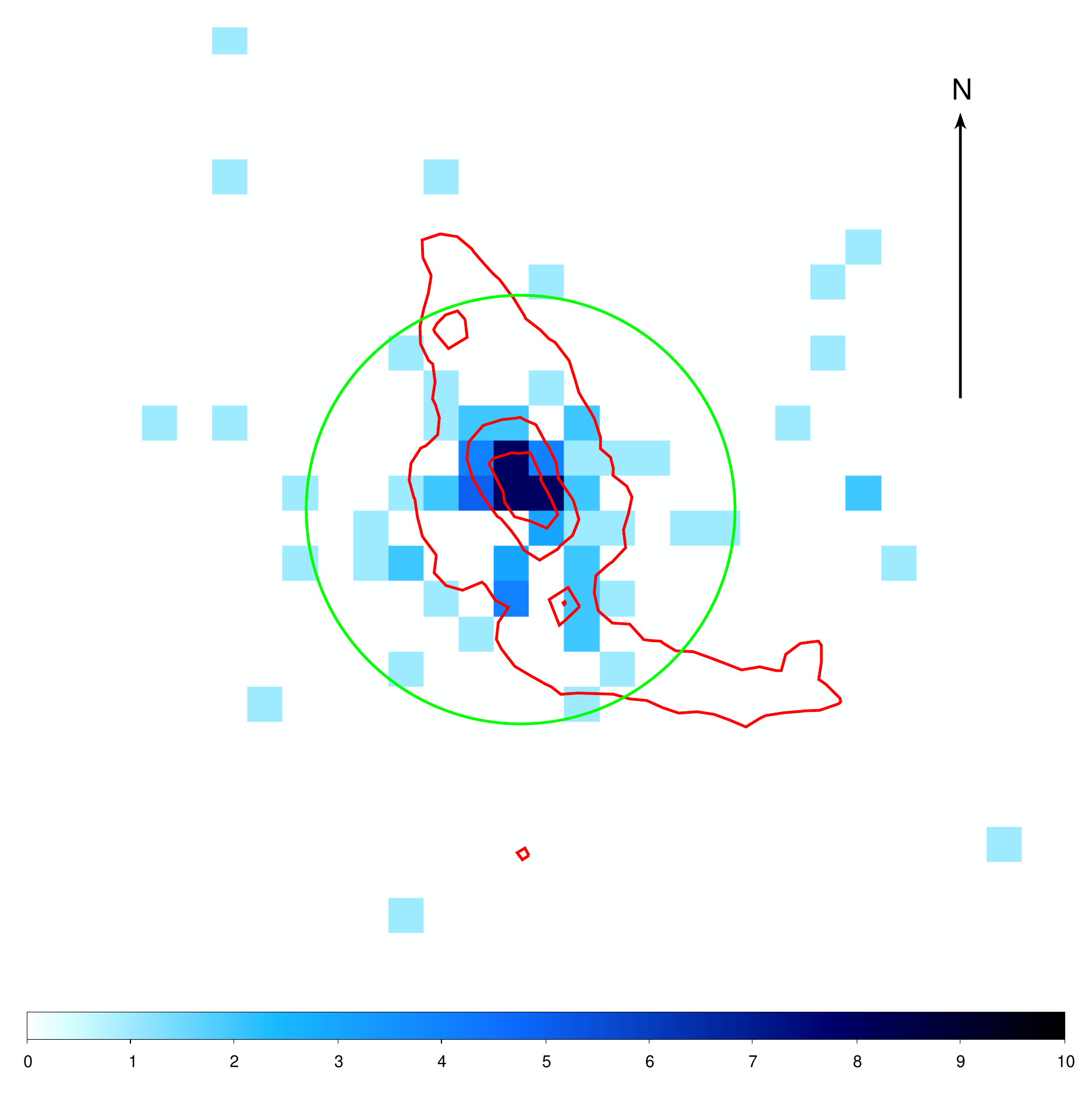}}
\centerline{\includegraphics[width=2.75in]{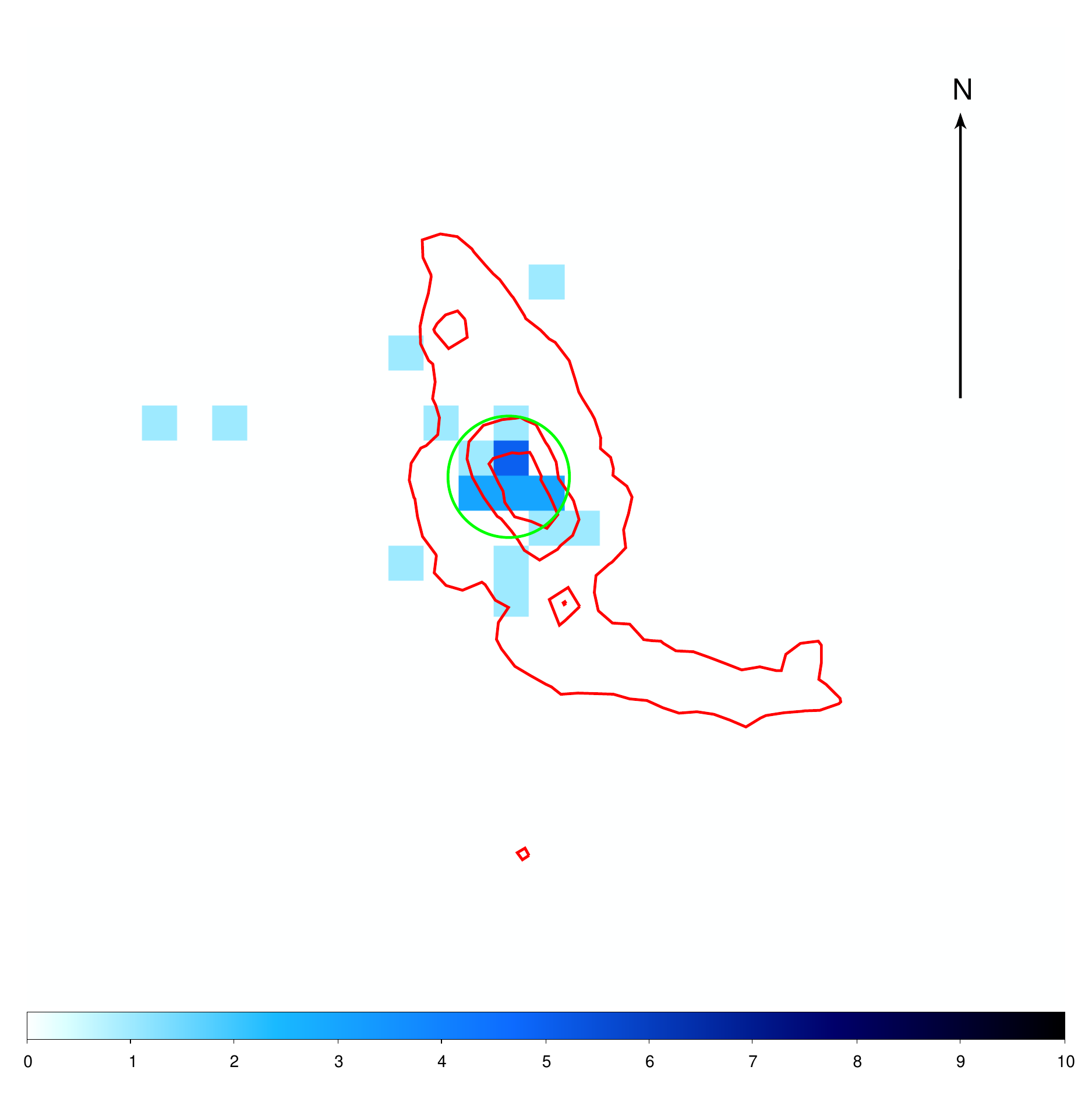}}
\caption{X-ray images of Tol 1247-232. Top -- full band (0.3--8~keV) X-ray image showing a bright point source near the peak of the optical emission and extended diffuse emission. The green circle shows the $3\arcsec$ radius extraction region used to obtain the spectrum of the whole galaxy. Bottom -- hard band (1.5--8~keV) X-ray image showing that the hard X-ray emission arises from a single unresolved source. The green circle shows the $0.85\arcsec$ radius extraction region used to obtain the spectrum of the unresolved source. The X-ray pixels are 0.492 arcseconds or 494~pc at the distance of Tol 1247-232. The red contours are from the optical image in Fig.~\ref{optical_image}. The black arrow points North and has a length of $4\arcsec$.}
\label{xray_images}
\end{figure}

To align the observations, we used wavdetect to search for sources on the S3 chip (where the target galaxy is located) and then searched for matches within $1.0\arcsec$ between X-rays sources and objects in the USNO B1 catalog, excluding the target galaxy. We shifted the astrometry of the individual observations in RA and DEC and then merged the two observations. We estimate that the astrometry is uncertain at the level of $0.5\arcsec$. X-ray images from the merged observations are shown in Fig.~\ref{xray_images}. There is diffuse X-ray emission spread across the central regions of the galaxy and about half of the emission is concentrated in a single unresolved source near the peak of the optical emission. The harder X-ray emission, at energies above 1.5~keV, is concentrated in the single unresolved source, near the center of the optical emission and consistent, within the astrometric uncertainties, with the positions of the star forming knots visible in Figure~\ref{optical_image}.

\begin{table}
\caption{X-ray spectral fits.}
\begin{tabular}{lcccc}
\hline
Spectrum  & $\Gamma$       & Flux                   & \multicolumn{2}{c}{Intrinsic Flux} \\ 
          &                & 0.5-8 keV              & 0.5-8 keV              &   0.5-2 keV  \\ \hline
Chandra   & 2.1$\pm$0.4 & $2.40^{+0.59}_{-0.47}$ & $2.67^{+0.60}_{-0.50}$ &  $1.42^{+0.37}_{-0.31}$ \\
Point     & 1.6$\pm$0.5 & $1.41^{+0.55}_{-0.40}$ & $1.63^{+0.60}_{-0.44}$ &  $0.57^{+0.18}_{-0.23}$ \\
XMM       & 2.2$\pm$0.3 & $4.41^{+1.15}_{-0.88}$ &                        & \\ \hline
\end{tabular} 
\newline
Note: Fluxes are in units of $10^{-14} \rm \, erg \, cm^2 \, s^{-1}$.
\label{xspec}
\end{table}

We extracted a spectrum from the whole galaxy, using a region with a radius of $3\arcsec$ shown in Figure~~\ref{xray_images}, and fitted it with an absorbed, redshifted powerlaw ($z = 0.048$) using the Cash statistic with the absorption column fixed to the Milky Way value of $6.6 \times 10^{20} \rm \, cm^{-2}$. The results, including the photon index ($\Gamma$), the observed flux (uncorrected for absorption), and the intrinsic flux (corrected for the assumed absorption), are given in Table~\ref{xspec} as ``Chandra''. Uncertainties on the spectral parameters are quoted at the 90\% confidence level. We also extracted a spectrum for a region with a radius of $0.85\arcsec$ centred on the hard X-ray emission in Figure~\ref{xray_images}. We fitted using the same model and the results are given in the table as ``Point''. In calculating the intrinsic point source flux, we corrected for the fraction of the point spread function outside the extraction region.

We re-analysed the XMM-Newton observations of Tol1247 obtained on 2005-06-22 \citep{Rosa2009}. We extracted a spectrum for the EPIC-PN using a circular region with a radius of $20\arcsec$ in order to encompass nearly all source counts and an annular background region with radii of $40\arcsec$ and $70\arcsec$ centred on the source. We used the same spectral model as for the Chandra data and fixed the absorption column to the Milky Way value. The results are in Table~\ref{xspec} as ``XMM''. Our measured flux is compatible with that reported by \citet{Rosa2009}, while our photon index is marginally softer, possibly due to the fact that we did not include any absorption intrinsic to Tol1247. We also extracted background-subtracted light curves and grouped the data into 1500~s time bins, obtaining at least 23 net counts in each of the 7 time bins. Following the timing analysis of \citet{Sutton2012}, we find a $3 \sigma$ upper limit on the fractional variability 0.26.

\section{Results and Discussion}\label{results}
\label{sec:results}
%\section{Discussion}\label{discussion}
%\label{sec:discussion}

The Chandra images, Fig.~\ref{xray_images}, show that Tol1247 produces soft emission spread across the central part of the galaxy and point-like emission with a luminosity of $(9 \pm 2) \times 10^{40} \rm \, erg \, s^{-1}$ and a hard spectrum with a photon index of $1.56 \pm 0.46$. Comparing the flux measured with {\it XMM-Newton} to the total flux measured with {\it Chandra} shows evidence for a decrease in luminosity of $1.0 \times 10^{41} \rm \, erg \, s^{-1}$ from 2005 to 2016. The variability is strong evidence for the presence of at least one highly luminous accreting source in Tol1247. The source region for XMM-Newton is larger than the galaxy, but it is unlikely that the extra flux originates outside the galaxy since the centroid measured with {\it XMM-Newton} is well centred on the galaxy. The {\it Chandra} point source may be the source of the excess {\it XMM-Newton} flux, in which case only a single accreting object is required. Otherwise, Tol1247 may contain a small number of luminous accreting sources that coincidentally decreased in luminosity together.
%The closest Chandra source is $28\arcsec$ away. 

The presence of one or more highly luminous accreting sources and the similarity of the X-ray emission from Tol1247 to that found from Haro 11 \citep{Prestwich2015}, the only other LyC-emitting galaxy that has been resolved in X-rays, may suggest a relation between accreting sources and LyC escape. Below, we discuss possible origins of the X-ray emission and the implications for the presence of accretion-powered outflows, the origin of the diffuse X-ray emission, and the relation  of accretion-powered sources to LyC escape.

\subsection{Nature of the Compact X-ray Source(s)}

The variable X-ray emission from Tol1247 may be due an active galactic nucleus (AGN). Tol1247 lies within the pure starforming region of the Baldwin-Phillips-Terlevich (BPT) diagram \citep{Baldwin1981} and there are no indications of an active nucleus at other wavelengths \citep{Leitet2013}. However, a minor AGN contribution cannot be excluded. The X-ray luminosity would place an AGN in the low-luminosity (LLAGN) regime where the accretion flow is sub-Eddington and thought to be radiatively inefficient and produce powerful outflows with mechanical energy comparable to or larger than the X-ray luminosity \citep{Ho2008}.

Another possible origin of the X-ray emission is from an ultraluminous or hyper-luminous X-ray source (ULX or HLX), for a review, see \citet{Kaaret2017}. ULXs are thought to be X-ray binaries that contain stellar-mass black holes or neutron stars in super-Eddington accretion states. HLXs may be super-Eddington accretors or intermediate mass black holes (IMBHs) with masses in the range $100-10^{5} M_{\sun}$. If the compact object is a stellar-mass black hole or a neutron star, then the accretion flow would be highly super-Eddington and strong outflows would be expected \citep{Kaaret2017}. If the compact object is an IMBH with a mass above $10^{4} M_{\sun}$, then the accretion flow would be sub-Eddington ($L_X/L_{\rm Edd} < 0.05$) and, hence, radiatively inefficient and the source of a powerful outflow, similar to that of a LLAGN. For a compact object mass near $10^{3} M_{\sun}$, the source could be in the near-Eddington X-ray thermal dominant state, where strong outflows are not observed.

The emission could, instead, arise from a small number of ULXs that happened to all decrease in flux from 2005 to 2016. The Chandra point source extraction region corresponds to 900~pc at the distance to Tol1247. This encompasses the whole of the central starburst in M82 with approximately 20 X-ray sources and emission dominated by two particularly bright objects, M82 X-1 and X-2. As noted above, ULXs are thought to be X-ray binaries in super-Eddington accretion states with strong outflows.

Repeated X-ray imaging of Tol1247 with {\it Chandra} could reveal whether the X-ray emission arises from a single source or multiple sources. Detection of strong X-ray variability, with rms of 10--20 per cent on timescales of 100s to 1000s of seconds as seen from several other HLXs \citep{Sutton2012}, would suggest that the source is in the hard X-ray state and that the compact object is an IMBH. Radio emission consistent with the fundamental plane relation between radio luminosity ($L_{\rm R}$), $L_X$, and black hole mass ($M_{\rm BH}$) for accreting black holes in the hard state \citep{Merloni2003} would positively identify the source as in the hard state and could place constraints on the black hole mass. We estimate that the radio flux at 5~GHz would be 14~$\mu$Jy for a $M_{\rm BH} = 10^{6} M_{\sun}$ and 2.4~$\mu$Jy for a $M_{\rm BH} = 10^{5} M_{\sun}$. Conversely, detection of X-ray pulsations would identify the compact object as a neutron star.
% The total 4.9~GHz flux from the galaxy is 2.3~mJy \citep{Rosa2007}, which may complicate detection of such a weak source.

One possible constraint on the nature of the X-ray emitter(s) is the age of star formation in the host galaxy. \citet{Buat2002} fitted the far ultraviolet spectrum of Tol1247 with spectral synthesis models and found the best fits for a recent starburst (1~Myr) or continuous star formation over at most 5~Myrs. The observed spectrum is bluer than any of the synthesis models. \citet{Rosa2007} concluded that Tol1247 contains a single, young star burst with an age of less than 4~Myr, based on the galaxy's high H$\beta$ equivalent width, high {\sc Oiii} to H$\beta$ ratio, shallow radio spectral index, and low ratio of 1.4~GHz versus H$\alpha$ flux. The optical results indicate that the stellar population is dominated by massive, young stars while the radio properties show that the starburst has not yet had time to produce a large number of type II supernovae that would power radio synchrotron emission. \citet{Puschnig2017} suggest that the galaxy has been forming stars for the past 30~Myr, but that the central region is dominated by very young stars with a mean age of 3~Myr. Therefore, the accreting object(s) responsible for the X-ray emission in Tol1247 likely evolved on a very rapid time scale, less than 4~Myr. 

Accretion from the interstellar medium onto a pre-existing massive black hole could occur on time scales as rapid as the gravitational infall of the gas leading to star formation.  Thus, a low-luminosity AGN intepretation for the X-ray emission is compatible with the young age of the star forming activity. The leading model for IMBH formation is dynamical interactions in star clusters. Simulations indicate that it is possible to form compact objects with masses above $1000 M_{\sun}$ in sufficiently massive and compact star clusters in less than 4~Myr \citep{Portegies2004}. This is compatible with the young age of the starburst. Indeed, a key requirement for runaway collisions of massive stars in a cluster is that they can reach the cluster center via dynamical interactions before exploding as supernovae. Conversely, formation of a high-mass X-ray binary is thought to occur after evolution of one member of the binary into the supergiant phase, requiring $\sim 12$~Myr \citep{Tauris2006}. This time scale is incompatible with the young age of the starburst in Tol1247. 
%The relatively high beaming required by the hyperluminous pulsar intepretation, by at least a factor of 7, would make it unlikely that HLX pulsars would be detected in both of the known, nearby LyC-emitting galaxies.

\subsection{Diffuse X-ray emission}

The Chandra images, Fig.~\ref{xray_images}, show that Tol1247 produces soft emission spread across the central part of the galaxy. Subtracting the aperture-corrected flux for the point source from the flux measured for the whole galaxy, we find that the absorption-corrected diffuse flux for the whole galaxy corresponds to an X-ray luminosity of $L_X = (4 \pm 2) \times 10^{40} \rm \, erg \, s^{-1}$ in the 0.5--2~keV band. The limited number of X-ray counts do not permit a detailed spectral analysis and the bolometric luminosity may be higher if the spectrum contains absorbed thermal emission at low temperatures.

Such soft diffuse emission is thought to be produced by energy injected into the interstellar medium by supernovae, stellar winds, and X-ray binaries, and, hence, proportional to the SFR. Different SFR indicators probe star formation on different times scales. As noted above, the starburst in Tol1247 is quite young. Using the SFR estimate from \citet{Rosa2007} of $47 \rm \, M_{\odot} \, yr^{-1}$ based on the H$\alpha$ luminosity and therefore appropriate for a young starburst and the relation between SFR and diffuse soft X-ray luminosity from \citet{Mineo2012b}, predicts $L_X = 4 \times 10^{40} \rm \, erg \, s^{-1}$ in the 0.5--2~keV band in agreement with the measured value. However, the $L_X-$SFR relation of \citet{Mineo2012b} was derived using IR and FUV as SFR indicators and may not be applicable for SFRs derived from H$\alpha$, also SFR estimates for Tol1247 vary significantly, e.g.\ \citet{Puschnig2017} find $35 \rm \, M_{\odot} \, yr^{-1}$ from H$\alpha$ and $96 \rm \, M_{\odot} \, yr^{-1}$
from 1.4~GHz radio emission.

The weak radio emission suggests that Tol1247 has not yet produced a large number of type II supernovae \citep{Rosa2007}. Hence, the diffuse emission must be powered by other means. \citet{Justham2012} suggest that X-ray binaries may dominate the energy input to the ISM for starburst ages less than 6~Myr and that the relative importance of X-ray binaries versus supernovae will be greatest for near solar metallicity galaxies with starburst masses near $10^8 \rm \, M_{\odot}$. Hence, Tol 1247-232, with a starburst mass of $1.3 \times 10^8 \rm \, M_{\odot}$ \citep{Rosa2007}, may be a good candidate for a galaxy in which feedback and heating of the ISM is dominated by X-ray binaries.
 %Due to small number of highly luminous X-ray binaries, their formation is subject to large stochastic fluctuations, so the rough agreement with the relation between SFR and $L_X$ measured using larger galaxies would be mostly fortuitous \citep{Justham2012}.}

\begin{figure}
\centerline{\includegraphics[width=2.75in]{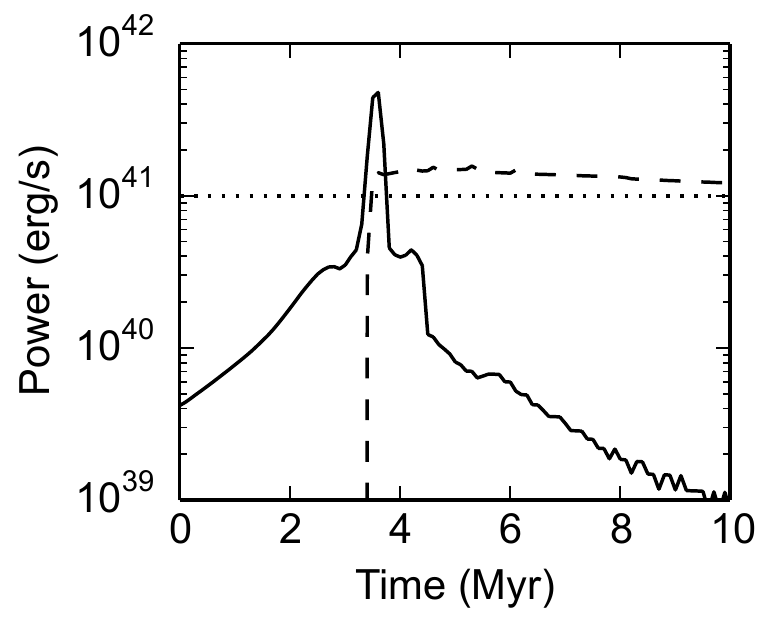}}
\caption{Mechanical power from an instantaneous burst with a total stellar mass of $10^7 \rm \, M_{\odot}$ versus age of the burst, following \citet{Prestwich2015}. The solid line is the power from stellar winds. The dashed line is the power from supernovae. The dotted line is at $10^{41} \rm \, erg \, s^{-1}$, representative of the level of accretion power present in Tol 1247-232.}
\label{power}
\end{figure}

\subsection{Accretion-powered outflows and Lyman-continuum escape}

The similarity of the X-ray emission from Tol1247 to that found from Haro 11, the only other LyC emitting galaxy that has been resolved in X-rays, may suggest a role for highly luminous X-ray sources in enabling LyC escape. Most of the possible interpretations of the source(s) of the X-ray emission observed from Tol1247 suggest the presence of at least one accretion-powered object that produces an X-ray luminosity near $10^{41} \rm \, erg \, s^{-1}$ and an outflow with a comparable mechanical energy. LyC photons may escape a galaxy via ionized or empty channels through the ISM leading out from the star-forming region. X-rays may photoionize the ISM. An outflow may either entrain matter, creating an empty channel, or deposit energy, ionizing the ISM. Any of these processes may help enable LyC escape.

To compare the stellar- versus accretion-powered mechanical energies, we used Starburst99 to model the mechanical power from an instantaneous star-forming burst with a Kroupa initial mass function with mass limits of 0.1 and 100~$M_{\odot}$, upper and lower exponents of 2.3 and 1.3, and a turnover mass of 0.5~$M_{\odot}$. Fig.~\ref{power} show the power for a total stellar mass of $10^7 \rm \, M_{\odot}$, similar to that measured in the bright knots B and C in Haro 11 \citep{Adamo2010} that appear similar to the knots in Tol1247, and at a metallicity of $Z = 0.004$, close to the metallicity of Tol1247 \citep{Terlevich1993}. The power output from the accreting X-ray source dominates until 3.3~Myr. Winds from Wolf-Rayet stars briefly dominate, reaching $5 \times 10^{41} \rm \, erg \, s^{-1}$. After 3.8~Myr, supernovae provide mechanical power similar to or somewhat larger than that of the accreting source. Hence, the accreting source may dominate the mechanical power at early times, depending on when it was formed, and provides a significant contribution at later times.

H$\alpha$ imaging of Tol1247 shows an arc with a diameter of $\sim 300$~pc near the location of the brighter star cluster \citep{Puschnig2017}, near the X-ray point-source. \citet{Puschnig2017} interpret this as evidence that feedback from the star formation in the star cluster produces outflows or a highly turbulent ionized medium that create at least one clear sightline for LyC to leak. They also find low extinction along the light of sight to the central star clusters. The young age of the central star clusters and deficit of radio emission argue against the feedback being due to supernova, but is consistent with feedback due to high-mass X-ray binaries.

It is not possible to draw robust conclusions from a sample of two objects, and an important question is whether some other factor drives LyC leakage and the presence of highly luminous X-ray sources is merely a by product. Many properties of Tol1247 and Haro 11 are similar, including the young age of their star clusters and their high central SFR surface density. \citet{Verhamme2017} have suggested that SFR surface density is the key determinant of LyC escape and that Lyman break analog galaxies (LBAs) and Green Pea galaxies (GPs) may be LyC emitters. Interestingly, both LBAs and GPs produce strong X-ray emission \citep{Basu2013,Brorby2017}. Conversely, there are HLX host galaxies that are not known to be LyC emitters. M82 is the closest example, however the strong absorption seen from the X-ray sources \citep{Kaaret2009} and the edge-on orientation of M82 are not favourable to observing LyC; it has been suggested that LyC escapes from M82 along other lines of sight \citep{Devine1999}. Inclination effects may be important if LyC escapes only along specific lines of sight. Searches for LyC emission or its proxies from HLX host galaxies and for X-ray emission from additional LyC emitters would be excellent tests of the relation between accretion-powered objects and LyC escape.

\section*{Acknowledgements}

We thank the referee for thoughtful comments that significantly improved the manuscript, particularly the discussion section. Support for this work was provided by the National Aeronautics and Space Administration through Chandra Award Number GO5-16081X issued by the Chandra X-ray Observatory Center, which is operated by the Smithsonian Astrophysical Observatory for and on behalf of the National Aeronautics Space Administration under contract NAS8-03060. We acknowledge the use of the HyperLeda database.  

%This research has made use of the NASA/IPAC Infrared Science Archive, which is operated by the Jet Propulsion Laboratory, California Institute of Technology, under contract with the National Aeronautics and Space Administration. 


%%%%%%%%%%%%%%%%%%%%%%%%%%%%%%%%%%%%%%%%%%%%%%%%%%

%%%%%%%%%%%%%%%%% APPENDICES %%%%%%%%%%%%%%%%%%%%%

%\appendix


%%%%%%%%%%%%%%%%%%%%%%%%%%%%%%%%%%%%%%%%%%%%%%%%%%

% Don't change these lines
\bsp	% typesetting comment
\label{lastpage}
\end{document}